# Ultraviolet Dual Comb Spectroscopy: A Roadmap

**VITTORIA SCHUSTER**[1,2], **CHANG LIU**[1,2,3], **ROBERT KLAS**[1,2,3], **PABLO DOMINGUEZ**[4], **JAN ROTHHARDT**[1,2,3,5], **JENS LIMPERT**[1,2,3,5], AND **BIRGITTA BERNHARDT**[1,2,6*]

[1]*Institute of Applied Physics, Friedrich Schiller University Jena, Albert Einstein Straße 6, 07745 Jena, Germany*
[2]*Abbe Center of Photonics, Friedrich Schiller University Jena, Albert Einstein Straße 6, 07745 Jena, Germany*
[3]*Helmholtz-Institute Jena, Fröbelstieg 3, 07743 Jena, Germany*
[4]*MenloSystems GmbH, Bunsenstraße 5, 82152 Planegg, Germany*
[5]*Fraunhofer Institute for Applied Optics and Precision Engineering, Albert-Einstein-Straße 7, 07745 Jena, Germany*
[6]*Institute of Experimental Physics & Institute of Materials Physics, Graz University of Technology, Petersgasse 16, 8010 Graz, Austria*
*\*bernhardt@tugraz.at*

**Abstract**

Dual Comb Spectroscopy proved its versatile capabilities in molecular fingerprinting in different spectral regions, but not yet in the ultraviolet (UV). Unlocking this spectral window would expand fingerprinting to the electronic energy structure of matter. This will access the prime triggers of photochemical reactions with unprecedented spectral resolution. In this research article, we discuss the milestones marking the way to the first UV dual comb spectrometer. We present experimental and simulated studies towards UV dual comb spectroscopy, directly applied to planned absorption measurements of formaldehyde (centered at 343 nm, 3.6 eV) and argon (80 nm, 16 eV). This will enable an unparalleled relative resolution of up to $10^{-9}$ – with a table-top UV source surpassing any synchrotron-linked spectrometer by at least two and any grating-based UV spectrometer by up to six orders of magnitude.

## 1. Introduction

High resolution spectroscopy is the most proliferate experimental technique used to identify the composition of matter. With lasers and white-light sources conveniently covering the visible and infrared spectral region, identification is based to an overwhelming share on detecting ro-vibrational transitions. Electronic energy levels in the ultraviolet (UV) spectral region are sparsely investigated due to the lack of direct accessibility with laser sources. However, unlocking this spectral window would facilitate fingerprinting the electronic energy structure of matter and thus unravel a fundamentally new range of quantum mechanical material properties.

Dual Comb Spectroscopy [1–4] (DCS) has proven its capabilities in molecular fingerprinting with its high spectral resolution, broad spectral coverage and short measurement times. Since the first demonstrations, various dual comb spectrometers have been realized in different spectral ranges and the number of their applications is still growing while their versatility keeps expanding. Most recent applications involve nonlinear spectroscopy [5], time-resolved and multidimensional studies [6,7] widening the field of applications dramatically. However, considering all DCS realizations so far, a striking vacancy remains in the ultraviolet spectral region (UV) that has so far not been conquered by DCS, mainly due to the lack of suitable laser sources. This article provides a roadmap paving the way for the expansion of dual comb spectroscopy towards the UV. It presents first important results towards its first realization covering the most important milestones: 1. high power dual comb source generation, 2. frequency-up conversion into the UV via high harmonic generation, 3. UV interferometry and 4. fast UV signal detection (see figure 1).

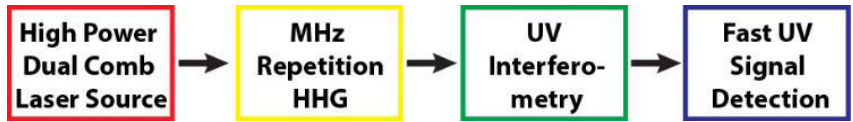

**Fig. 1** Roadmap for UV dual comb spectroscopy highlighting the four most important milestones for its realization: 1. high power dual comb source generation, 2. frequency up conversion into the UV via high harmonic generation (HHG) at MHz repetition rates, 3. UV interferometry and 4. fast UV signal detection.

## 2. Dual comb spectroscopy

For a detailed description of the DCS principle, it is referred to recent comprehensive reviews on the topic [3,4]. Here we only remind of the main features and characteristic variables that become especially relevant for the discussion of the applicability of DCS in the UV.

DCS realizes the concept of Fourier transform spectroscopy (FTS) by superimposing the output of two broadband laser frequency combs with slightly detuned repetition rates $f_{rep}$ and $f_{rep}+ \delta$, see Fig. 2. The detuning $\delta$ causes the pulse trains of the two laser sources scanning over each other in time. These pulse pairs with sweeping time delay mimick the mechanically introduced path length difference involved in conventional interferometric Fourier transform spectrometers. The sweeping happens orders of magnitude faster and in a wider time delay range than any mechanical setup would permit. In the frequency domain, this results in a down conversion of the optical spectrum exhibiting a full width of $\Delta \nu$ (typically >> 10 THz) into the radio frequency domain of width $\Delta \nu/a$, where $a = f_{rep}/\delta$ is the down conversion factor.

In the time-domain, the interference signal of the two pulse trains, the interferogram, reproduces itself with a period of $T = 1/\delta$. Typical values are $\delta$ = 50-1000 Hz and $T$ = 1-20 ms, respectively. The minimum required measurement time, however, is much smaller than the interferogram's period and depends on the sample and the down conversion factor: The interrogating frequency comb excites the sample at regular time intervals and the subsequent free induction decay is sampled with the second comb, detuned in its repetition frequency. For obtaining the absorption information of the sample, it is sufficient to perform the Fourier transformation of the modulated section in the interferogram, i. e. the ringing dipole that arises from the induction decay. The down conversion in the frequency domain corresponds to a stretching in the time domain, allowing to observe the actual *effective* time scale of the induction decay on a slower *laboratory* time scale. Typical minimum measurement times (laboratory time) are on the order of tens of μs while the "real" atomic or molecular response lasts typically nano- or even picoseconds (effective time). Those short "single shot" measurements can be performed without active stabilization of the combs. Longer measurement times that span many periods of the interferogram increase the signal to noise ratio and the spectral resolution to the kHz level [8]. For that, long-time mutual coherence of the two comb sources is required. If this is achieved, the molecular absorption features of the sample can be resolved by the repetition frequency of the frequency combs, sometimes referred to as sampling point spacing. The repetition rates can range from several tens of MHz (fiber or solid state frequency combs) to several GHz (micro resonators). Even higher resolution can be achieved by interleaving subsequently recorded spectra with shifted carrier envelope offset (CEO) frequencies.

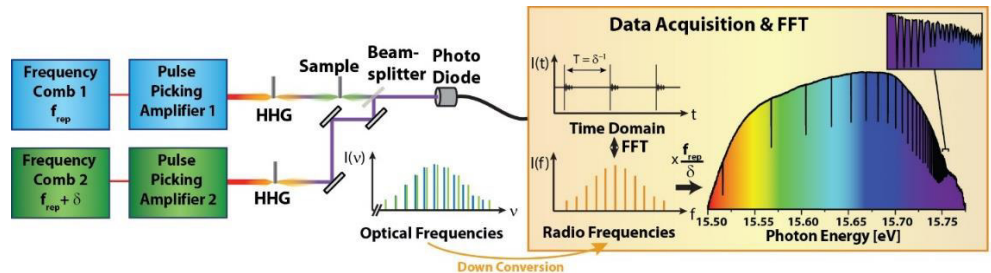

**Fig. 2** Concept of dual comb spectroscopy and its extension to applications in the (extreme) ultraviolet spectral region via frequency up-conversion (high harmonic generation, HHG). The outputs of two frequency combs with slightly different repetition rates are intensified with pulse picking amplifiers and frequency-up-converted into the ultraviolet spectral range were many molecular and atomic samples exhibit strong and congested absorption features. In this sketch, after one comb interacted with the sample, the two beams are overlapped with a beam splitter and the interference is detected with a photo diode. Other possibilities of interference of the UV frequency comb outputs are also considered in this work (see section 3.3.3). The Fourier transformation of the time-domain interferogram reveals the absorption spectrum of the sample (depicted here: argon Rydberg series), with high spectral resolution down to the µeV level (corresponding to a relative resolution of up to $10^{-9}$).

Since the most recent review article covering dual comb spectroscopy [4], further extensions of dual comb spectroscopy in the THz, infrared and visible spectral region have been accomplished, see for example [9–15]. However, the near ultraviolet still has only been touched barely and indirectly via second harmonic generation and two photon excitation [16,17]. Although the conversion of frequency combs to the UV via different higher order frequency-up conversion approaches such as high harmonic generation (HHG, [18]) has repeatedly been demonstrated (see [19,20] and references therein), this spectral region has not yet been conquered by dual comb spectroscopy.

## 3. Dual comb spectroscopy in the ultraviolet spectral region

### 3.1 Motivation

A variety of compound families exhibit congested multi-line spectra in the ultraviolet range due to the $\sigma \to \sigma^*$ and $\pi \to \pi^*$ excited state transitions in many alkanes, halogens, dienes, aromatics and carbonyls. Especially, the air pollutants nitrous oxides, the majority of the greenhouse gases and astrophysically relevant gases have striking spectral signatures in the UV with large absorption cross sections reaching hundreds of Megabarn (1 Mb = $10^{-22}$ m²) [21–23]. Detailed knowledge of their spectral features is essential as high-resolution benchmark data help to improve the understanding of the processes in those elemental gases, help to determine their detailed composition and help to render more precise ab-initio line shape parameter calculations for astrophysics and atmospheric sciences [24,25]. Laboratory measurements as counterpart to environmental monitoring data or satellite mission data have been carried out at synchrotron sources for more than 40 years but an extensive database is still inexistent as only very few synchrotron beamlines provide a sufficient spectral resolution (relative resolution up to < $10^{-6}$, 100 μeV, < 10 GHz). The relative spectral resolution of state of the art literature data ranges from $10^{-4}$ to $10^{-6}$, often insufficient for accurate determination of transition linewidths. This is known to result in the under-estimation of absolute photo absorption cross sections [26]. UV Dual Comb Spectroscopy could improve our state-of-the-art knowledge of the UV absorption behavior of those gases by up to three orders of magnitude as it is mainly limited by the repetition rate of the frequency combs used. This refinement in spectral resolution would for example allow for the first time a spectroscopic analysis of $CH_3I$ based on the framework of a multichannel quantum defect approach instead of applying the simple Rydberg formula [22]. A recent feasibility study of UV DCS shows its potential importance for atmospheric trace gas applications [27]. Broadband high resolution absorption measurements of gasses of atmospheric interest could for example shed light onto photodissociation processes induced by solar (UV) radiation and onto the onset of aerosol nucleation.

Another aspect is the extremely large spectral coverage that comes along with the frequency up-conversion into the UV (see upcoming section): each harmonic arising from HHG covers far more than 100 THz outperforming any MIR dual comb spectrometer at least by a factor of two [28].

Very recently, using three frequency combs [7] or a delay line [6] in the infrared region (1550 nm), first studies witness the applicability of DCS to time-resolved experiments. The combination of UV dual comb and pump-probe spectroscopy could revolutionize ultrafast spectroscopy as it combines high temporal resolution on the fs time scale with μeV spectral resolution. UV-DCS is not limited to the investigation of gaseous samples. The absorption properties of liquid and solid state samples can also be explored with DCS. This has already been realized efficiently via near-infrared (NIR) optical coherence tomography for the investigation of different kinds of surfaces [29]. Exploiting the wavelength dependence of the diffraction limit, XUV dual comb spectroscopy has the potential to qualify for the detailed investigation of three dimensional structures on the nanometer scale, similar to [30] but with the capacity of a higher photon flux by abandoning the necessity of XUV gratings.

*3.2 The experimental setup*

For dual comb spectroscopy in the ultraviolet spectral region, the basic concept of superimposing two frequency combs with slightly different repetition rate remains the same. However, the UV frequency combs have to be realized via frequency up-conversion techniques like high harmonic generation (HHG, see fig. 2) because there are no frequency combs directly generating the UV radiation. For this, we propose the following concept: The output power of two detuned frequency comb oscillators (emitting in the infrared region) is enhanced with two pulse picking chirped pulse amplifiers. The pulse pickers enable very flexible peak intensities for low harmonic generation in the near ultraviolet and high harmonic generation in the extreme ultraviolet spectral region. Depending on the spectral region of interest, different concepts of superimposing the two UV frequency comb outputs might be favorable. In fig. 2, a direct superposition of the UV beams is depicted with a beam splitter. For high photon energies in the XUV, this will result in a very limited spectral bandwidth and low transmission through the combining beam splitter. The following sections present our concept in more detail.

*3.3 The four milestones towards UV dual comb spectroscopy*

Extending frequency combs and the interferometric principle of dual comb spectroscopy into the UV brings along a number of challenges. Where needed, a differentiation between the vacuum ultraviolet (n, 3.1 - 12.4 eV, 100 - 400 nm) and the extreme ultraviolet range (XUV, 12.4 - 124 eV, 10 – 100 nm) is applied. For simplicity, the following discussion will consider exemplarily two specific implementations at 340 nm (or 3.6 eV, for the VUV) and at ~ 80 nm (or 15.6 eV, for the XUV). Around 340 nm, the photodissociation of $NO_2$ and the dense absorption spectrum of $CH_3I$ could be studied for example. This could result in a more accurate absorption cross section of $CH_3I$, which remains underestimated due to the insufficient spectrometer resolution so far. High resolution spectroscopy around 15.6 eV enables the detailed study of the argon Rydberg series converging against the first ionization energy (see fig. 2) that has so far been studied in the highest detail with millimeter wave spectroscopy [31]. UV DCS promises a comparable ultimate spectral resolution level while the investigation could involve also smaller quantum numbers than before, that means $n$ < 30.

*3.3.1    High power dual comb laser source*

Frequency up-conversion of a (NIR) frequency comb via HHG has been established for table-top XUV frequency comb generation. Due to the high nonlinearity of this process, high peak intensities on the order of $I_{peak} \sim \frac{2}{\pi w_0^2} \frac{P_{ave}}{\tau f_{rep}} \gg 10^{13} \frac{W}{cm^2}$ are required, with $P_{ave}$ being the average power, $\tau$ the pulse duration and $w_0$ the beam radius at the focus in the interaction region. To achieve the critical intensity for HHG, high average powers, small focus geometry, short pulse durations or/and small repetition rates can be applied. So far, few cycle laser systems with repetition rates on the order of 1 to 10 kHz are dominant for table-top XUV sources involved in attosecond science. For XUV-DCS, the aspects of the extremely high optical frequencies and the broad spectral coverage have to be taken into account. Considering the argon Rydberg series (fig. 2), it is sufficient to select one harmonic that covers the spectral area around the first ionization energy at 15.6 eV. This applies for example for the 13$^{th}$ harmonic of an Ytterbium fiber based laser system operating at 1030 nm (or 1.2 eV) [32]. The spectral width of this single harmonic is typically about 155 THz. In order to down convert this broad bandwidth without aliasing, the repetition rates of two 100 kHz laser systems would need to be detuned by only 32 µHz. We succeeded in demonstrating such a small repetition frequency detuning by standard piezo actuator technology, however, the associated period in the interferogram of more than eight hours makes this parameter configuration useless. A more practicable period in the interferogram and a decent noise performance set the minimum repetition rate of XUV-DCS to 10 MHz. With this, the detuning results in ~ 0.3 Hz and the interferogram periodicity to 3 s. Another option would be to reduce the spectral coverage of the UV sources because this enables higher repetition rate detuning without aliasing and thus a small periodicity in the interferogram.

*The phase noise of the amplifier system*

The carrier phase noise of the laser frequency comb sources critically influences the achievable sensitivity of the spectrometer. The phase noise and the achievable signal-to-noise performance of (infrared) dual comb spectroscopy have been studied in detail by N. Newbury et al. [33]. The pulse picking fiber amplifier does not only amplify the seed signal and its noise but additionally introduces excess noise mainly due to the noise of the pump(s) used in the amplifier, the pulse picking AOM that optionally decreases the comb`s fundamental repetition frequency and environmental effects from the laboratory (e.g. acoustics). In the presented case, our fiber frequency comb sources have a fundamental repetition rate of 80 MHz. To optimize the parameters for DCS (see above) and HHG (see section 3.3.2), the repetition rate needs to be decreased, realized via a pulse picker, both for XUV-DCS and for the experiments presented in this work.

In order to characterize the noise performance of our amplifier system, we conducted measurements with a balanced cross correlator, BCC-PD by Menlosystems [34], first with our two Yb fiber frequency combs alone, then with one of the combs amplified. Please see the supplementary information for details on the measurement settings. Figure 3 summarizes the behavior of the amplifier, panel a showing different pump current settings (average output powers of 25 W, 40 W and 70 W) and panel b presenting different pulse picking conditions (f$_{rep}$ = 4 MHz, 10 MHz, and 80 MHz). Panel b shows that decreasing the repetition rate adds the phase noise above the picking frequency into the baseband by aliasing [35]. As a result, the AOM (pulse picker) seems to add significant noise to the amplifier for high picking ratios and hence the repetition rate should not be chosen too low. Our phase noise measurements prove that the amplifier noise performance at a repetition rate level of 10 MHz is adding negligible noise contribution (< 10 fs) to the overall performance of the laser system.

Further improvement of the amplifier could involve the implementation of a referenced instead of free-running AOM as pulse picker. With this, phase-stable frequency comb operation with a similar pulse picking system has most recently been achieved [36]. There, a similar effect of the AOM on the carrier envelope phase has been shown: while the integrated phase noise of the oscillator and the amplifier both operating at a repetition rate of 63.7 MHz was 56 mrad and 160 mrad, respectively, the phase noise was higher in case of pulse picking operation (360 mrad at $f_{rep}$ = 100 kHz). The phase noise performance could however be further improved for example by implementing acousto-optic frequency shifter in a feed-forward configuration after the amplifiers [37].

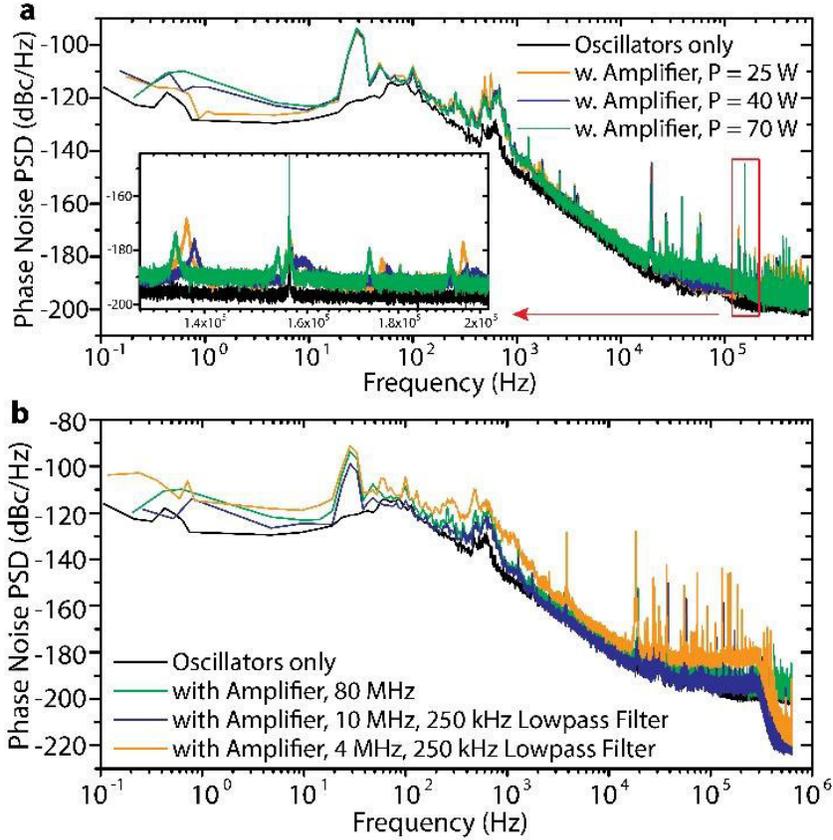

**Fig. 3** Phase Noise Power Spectral Density (PSD) derived from balanced cross correlation measurements of the two frequency comb oscillators alone (black, depicted in both panels) and with one fiber amplifier included. **a** PSD for different output powers: 25W (orange), 40 W (violet) and 70 W (green), respectively. The inset depicts a close-up ranging from 130 to 200 kHz and showing how the noise peaks shift with the pump current. **b** PSD for different repetition rates: 80 MHz (green), 10 MHz (violet) and 4 MHz (orange). See text for more details. Please see supplementary information for more details on the measurement settings and the analysis of the noise performance. See Supplement 1 for integrated phase noise plots.

### 3.3.2 Frequency up-conversion into the UV at MHz repetition rates

While reaching the VUV spectral range can be accomplished for example by third harmonic generation with an Ti:Sapphire or Yb fiber based laser frequency comb [38], the generation of XUV frequency combs is so far only established via HHG. With pulse repetition frequencies on the order of MHz, it is challenging to achieve the required peak intensities on the $10^{13} – 10^{14}$ W/cm$^2$ level but in recent years, several attempts based on both Ti:Sapphire and Ytterbium fiber based amplifier systems, seeding so-called enhancement cavities for upscaling the pulse energies, have been successful (see for example [19,20,39]).

However, there are still only strikingly few spectroscopic experiments carried out with enhancement-cavity-based XUV frequency comb sources because their continuous operation is challenging [40–42]. For XUV-DCS, the requirement of two simultaneously stabilized enhancement cavities turn the operation into hardly manageable. Efforts towards simplified setups involving enhancement cavities exist but XUV dual comb spectroscopy could not yet be performed [43]. In recent years, the ytterbium fiber based high power amplifier technology has advanced such that single pass HHG with MHz repetition frequencies and decent photon fluxes of $10^{12}$ - $10^{13}$ photons/s are achieved, yielding several tens of μW per harmonic across a spectral region of 20 – 30 eV (40 - 60 nm) with a conversion efficiency of $10^{-6}$ [32]. With this development, XUV-DCS comes within reach as the involved single pass geometry circumvents several problems that arise from intra-cavity dispersion, nonlinearities and plasma dynamics. For repetition rates below 20 MHz, the steady-state ionization fraction becomes negligible because the time interval between subsequent pulses gets larger than the plasma transit time [20].

The demanding parameters for XUV-DCS involve cutting edge laser technology that is currently under development for demanding spectroscopic applications. While these efforts are ongoing, we present here our pre-studies addressing important issues such as efficient interferometry in the XUV and the coherence of dual pulse XUV generation, both especially relevant for applications below 200 nm (> 6 eV).

*The phase noise of frequency up-conversion*

While for low order frequency multiplication, the phase noise power spectral density around a carrier increases quadratically with the harmonic order [44], the phase noise behavior of HHG can be more complex. In addition to the phase noise of the driving laser being transferred to the XUV, there can be other contributions like amplitude-phase coupling during the HHG process. A fluctuation of the target gas density within the time scale of the pulse-to-pulse time separation (the inverse of the repetition rate) can affect the pulse-to-pulse phase shift of both the driving beam and the harmonics beam. Plasma density fluctuations can also introduce phase noise into the driving lasers' and the harmonics' laser electric fields because of group velocity mismatch between laser driver and XUV light.

The phase noise of XUV frequency combs has been experimentally studied in the past, both for enhancement cavity-assisted and single-pass HHG [45,46]. It has been shown that high harmonic generation is highly phase coherent owing to the rigorous one-to-one mapping between moment of ionization launching the three-step-process [47], instantaneous field amplitude and emitted photon energy. An experiment with cavity-assisted HHG and an active interferometer stabilization could demonstrate that the quadratic evolution of the phase noise with rising harmonic order applicable to low order frequency multiplication is not "HHG physics" [45], but instead could show the high coherence of HHG by maintaining the linewidth on the 10 mHz level up to the $17^{th}$ order (60 nm or 20 eV). This high coherence is rooted in the highly non-perturbative nature of HHG that directly phase-locks the event of electronic wavepacket launching into the ionization continuum, the subsequent acceleration and recombination cycle to the fundamental laser electric field. As a result, the HHG process does not introduce timing jitter (or equivalently, phase noise bandwidth) and the coherently generated XUV frequency comb follows instantaneously its fundamental frequency comb bursts. One direct effect of this high coherence is the observation of sub-cycle features in time-domain attosecond experiments even with free running few-cycle laser systems. Here, the carrier envelope phase, that affects the offset frequency of the frequency comb, is not stabilized but the recorded absorption transients in the XUV directly map the phase of the fundamental laser's electric field [48]. Other experiments performing Ramsey-type spectroscopy using twin-pulse amplification and single-pass HHG frequency-up conversion proof that the XUV-DCS scheme is robust under a typical amplified frequency comb's repetition rate jitter performance [49].

The pulse picking Yb amplifiers that we use can reach 160 mrad phase noise with CEO stabilization (integration bandwidth: 10 Hz to 21 MHz) [36]. With active stabilization of the comb parameters and the optical path, the phase noise can be maintained in the XUV similar to [45]. Additional phase noise might arise from intensity fluctuations of the laser/amplifier source affecting the timing of HHG emission. For our systems we find rms deviations < 0.8 % for the average power (20 minutes measurement time) [50] allowing long term measurements, averaging and signal processing. These parameters will allow XUV DCS to yield a relative spectral resolution of $10^{-9}$.

### 3.3.3 UV interferometry

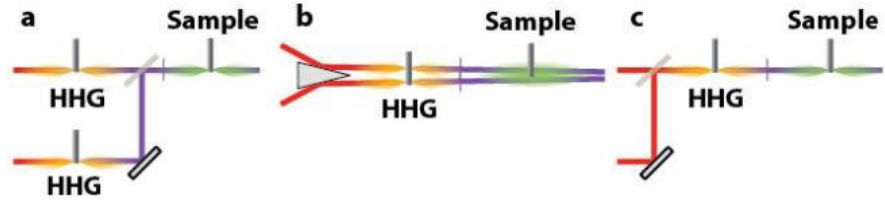

**Fig. 4** Three setups for (X)UV interferometry. **a** Superposition in the XUV, **b** quasi-collinear HHG, **c** superposition in the NIR. See text in section 3.3.3 for details.

The superposition of the two UV frequency combs for their interference on the detector without losing too much UV photon flux requires consideration. The two UV beams can be overlapped before or after sample interaction. In case of only one comb interrogating the sample (as depicted in figure 2), additionally to the absorption, also the dispersion of the sample can be measured by a simple photo diode. After Fourier transforming the time dependent interferogram, the sample's absorption (and dispersion) spectrum is retrieved.

For many years, the wavelength range for interferometric setups like Fourier transform spectroscopy was limited to $\lambda > 178$ nm because of the silica beamsplitters generally used in the spectrometer setup. The development of $MgF_2$ beamsplitters shifted the wavelength cutoff for FTS to 140 nm [51]. This wavelength limit could be shifted further down to 40 nm by switching from amplitude division (e.g. conventional beam splitters) to wave front division [52]. We examined different concepts of XUV interferometry that pave the way for DCS applications even deeper in the XUV (< 40 nm).

We could identify three promising configurations for an efficient superposition of the XUV beams:

**a** Performing HHG in two well separated gas jets and subsequent superposition of the XUV beams (fig. 4a). With this, the HHG parameters can be separately optimized for the two NIR combs. However, rather high losses will occur while superimposing the two XUV beams due to the large dispersion and material absorption coefficients in the XUV, and the spectral bandwidth is limited. An XUV beam splitter based on an ultra-thin $Si_3N_4$ membrane coated with a Si/Mo layer structure could achieve a reflectivity of 14.2 % and a transmission of 15.2 % at $E_c \sim 90$ eV or $\lambda_c \sim 14$ nm [53]. This means rather high losses for the XUV photon flux that would be even further increased for ultra-broadband applications. Hence, we consider also the following alternatives that promise to involve smaller losses:

**b** HHG in two gas jets located very closely side-by-side resulting in two XUV beams co-propagating under a small angle such that they interfere in the far field on the XUV photo diode (fig. 4b). This has the advantages of circumventing any transmissive optics and enabling individual adjustment of the HHG parameters for both beams. However, the wave fronts are not completely overlapping resulting in interference fringes and decreased signal-to-noise ratio of the interferogram. Despite slightly detrimental to the fringe contrast in the detector plane, this geometry has proven successful for two-dimensional and coherent diffraction imaging in the XUV [54–56]. For UV-DCS, focusing with a grazing incidence toroidal mirror could improve the signal-to-noise ratio for the detection.

**c** Superposition of the two NIR beams and subsequent HHG in a single gas jet (fig. 4c). This has the advantage that the beam guidance and superposition is efficiently possible with well-developed NIR optics. As a result of the joint HHG in a single gas jet, disturbing interference and plasma effects can occur due to the two NIR pulse trains influencing each other by ionizing the gas jet such that the second pulse of the pulse pair will be converted less efficiently than the first one. We examined this latter scenario and how those effects can be compensated by balancing the power of the two XUV beams (see upcoming paragraph and figure 5). When the two pulses temporally overlap in the gas jet, the NIR interference effects can be neglected within the frame of signal post processing for certain applications, especially when the induced induction decay in the molecules is much longer than the NIR pulse duration. Simulation results on the post processing follow at the end of this section.

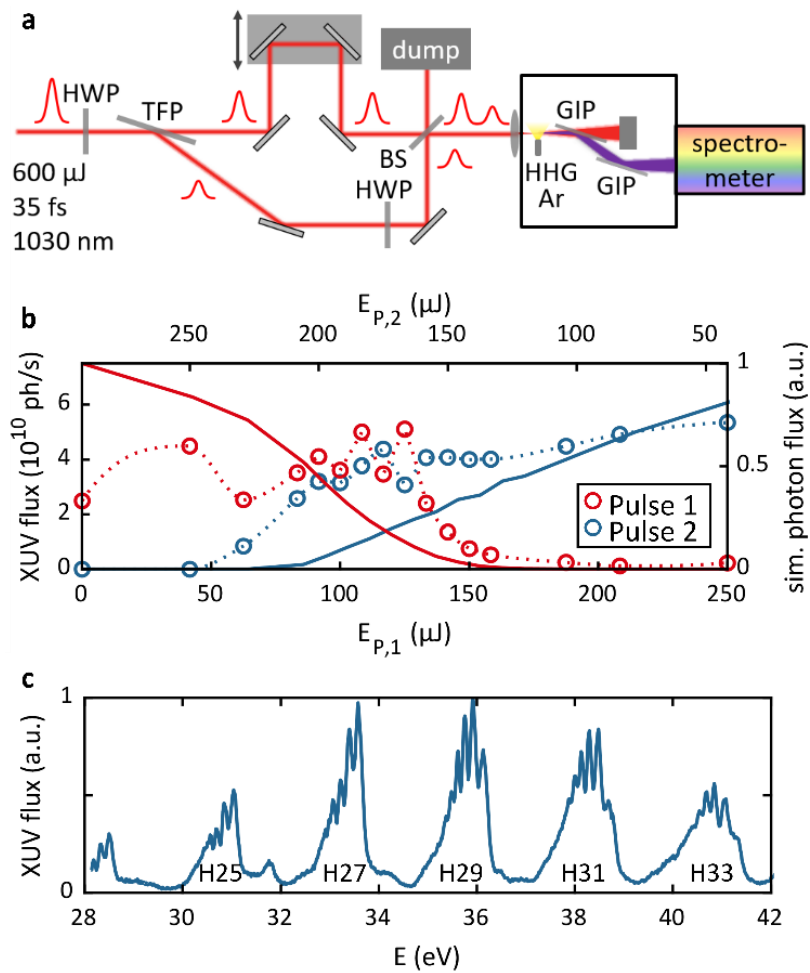

**Fig. 5** Dual Pulse Generation in the XUV. **a** Setup of the XUV interferometer with adjustable ratio of the pulse energies in the two interferometer arms. HWP Half wave plate, TFP Thin film polarizer, BS beam splitter, GIP Grazing incidence plate. **b** Adjustment of the NIR pulse energies for aiming at an equal XUV photon flux for the pulses arriving first and second in the gas target. The measured (left y-axis) and simulated (right y-axis) XUV photon flux is plotted over the NIR pulse energy. As the sum of the pulse energies remains constant, the increase of the energy of the first NIR pulse $E_{P,1}$ (bottom x-axis) corresponds to a decrease of the second pulse's energy $E_{P,2}$ (top x-axis). All measurements are recorded at a fixed time delay of 870 fs for the 33$^{rd}$ harmonic. The contribution of the first NIR pulse to the generated XUV flux is visualized in red, the flux from the second pulse in blue. Circles correspond to measured values (with dotted lines as guide to the eye), solid lines to simulations (see Supplementary for details). **c** Interference fringes observed in the XUV at a time delay of 18 fs. The orders of the recorded harmonics are marked in the bottom of the plot. The normalized XUV spectrum is given in dependence of the photon energy.

In a pre-study, an XUV Michelson interferometer was realized, seeded with one amplified frequency comb (fig. 5a). With this, two pulse trains with variable time delay similar to the DCS configuration in fig. 4c can be imitated and the effects occurring in a single HHG gas jet can be studied. The used laser system is similar to the one presented in [57]. The system was optimized for pulse energies of 600 µJ (average power 7 W, repetition rate 12 kHz, pulse duration < 40 fs). This allows for short integration times of the XUV CCD camera of 0.1 s, favorable to study the coherence properties of the configuration for twin-pulse-XUV generation in a single gas target. For this, the output of an amplified frequency comb is split by a broadband thin film polarizer (TFP). The splitting ratio can be adjusted with a half wave plate (HWP) in front of the TFP. One of the two interferometer arms includes a referenced delay stage for adjusting the time delay of the pulses in the HHG gas jet, similar to two overlapped frequency combs with detuned repetition rates. The two beams are overlapped with a beam splitter (BS) and focused into an argon gas jet for HHG in a vacuum chamber.

After separating the XUV from the fundamental with two grazing incidence plates (GIP) [58] and two Al filters (thickness: 200 nm), the generated XUV radiation is detected by an XUV flat-field grating spectrometer equipped with a CCD camera.

The target gas is ionized by the pulse arriving first yielding a decreased number of neutral gas atoms available for HHG to the second pulse and affecting its phase matching conditions. If the two pulses were of equal strength, the XUV comb generated by the pulses arriving second would have a lower XUV flux than the XUV comb generated by the "first" comb that hits the entirely neutral gas. Figure 5b shows that a balanced XUV photon flux can be reached for these laser parameters and beam geometry by increasing the pulse energy of the second pulse to about 1.5 as much as the pulse arriving first in the jet. Our simulations on twin pulse HHG predicted this behavior (see solid lines in fig. 5b and Supplementary for details). This adjustment of the pulse energies has been recorded at a constant time delay of 870 fs where the NIR pulses do not overlap temporally.

For small time delays similar to and smaller than the NIR pulse duration, interference between the two NIR pulse trains occurs, resulting in transient drops and peaks of the XUV output recurrent with a period that is corresponding to the laser's fundamental wavelength (see **Visualization 1**). However, this NIR interference has no negative effect onto the coherence of the HHG process in the single gas jet since we observe distinct interference fringes in the XUV spectrum. Such a spectrum is shown in figure 5c, recorded at a time delay of 18 fs. The limited contrast of the fringes is due to the vibrations occurring in the interferometer during the measurement (integration time: 0.1 s) and the rather low resolution of the grating spectrometer (100 meV).

For performing FTS or DCS in the XUV in the configuration of XUV-twin-pulse-generation in a single jet (figs. 4c, 5a), the interferogram does not only exhibit the XUV interference with the sample's induction decay but also the periodic NIR interference described above will be superimposed on the time trace. This can hamper the signal to noise ratio of the sample response that can be extracted from the interferogram. On the other hand, given the short duration of the NIR pulses, their interference happens on a very short time scale when compared to the sample lifetimes: For example, the life times of (atomic) Rydberg states are on the order of ps or even ns, and hence 100 to 100.000 times longer than the 35 fs short NIR laser pulses. As a result, the induced dipole oscillates at least one hundred times longer than the potentially disturbing NIR interference effect lasts. Hence, an adjusted apodization window neglecting the NIR interferences seems possible with losing only negligible signal strength of the molecular response - in this case of narrow absorption lines and long ringing dipole oscillations. For broad spectral features (linewidths approaching 1 THz), the proposed apodization treatment does not qualify because too much signal would be lost. However, typical UV linewidths of gaseous samples are on the 10-100 GHz level and hence the proposed apodization treatment is largely applicable.

While a fast and efficient XUV signal detection required for this experiment is under development (see 3.3.4), the influence of such an apodization window onto the retrievable absorption spectrum has already been simulated. This is especially applied to the case of XUV Fourier transform spectroscopy of argon Rydberg states converging against the $^2P_{3/2}$ ionization limit. For this, a high resolution argon XUV absorption spectrum is used [59] and applied to the 13$^{th}$ harmonic of our laser system (see fig 6 and Supplementary for more details). In order to retrieve the corresponding (simulated) interferogram that would arise from an XUV-FTS measurement in argon, the Fourier transform of this spectrum is calculated (fig. 6a).

By applying different apodization windows with different starting times and window slopes to the interferogram, it can be considered how much of the molecular response will be left by analyzing the SNR of the resulting absorption spectra. Figure 6a shows the full interferogram and an interferogram with an apodization window applied ($t_{AP}$ = 1000 fs). Additionally, the auto correlation trace of our NIR amplifier output is shown (red line). Figure 6b compares the absorption features after Fourier transforming the interferograms depicted in fig. 6a, respectively. A close-up to one Rydberg state (n=16, arbitrarily chosen) shows the influence of the apodization to the line shape of the resonances (fig. 6c). The line shape parameters change on the µeV level with regard to the line width and on the sub-µeV level with regard to the resonance center positions. The rather generous cut of $t_{AP}$ = 1000 fs excludes also a small NIR side pulse at 500 fs (see autocorrelation trace in fig 6a). The amplitude results in 91 % of the incident value. With this, it can be concluded that applying apodization windows to the interferogram is a promising option to neglect potentially disturbing NIR interference effects provided the NIR pulse duration is small compared to the sample response time (state lifetimes).

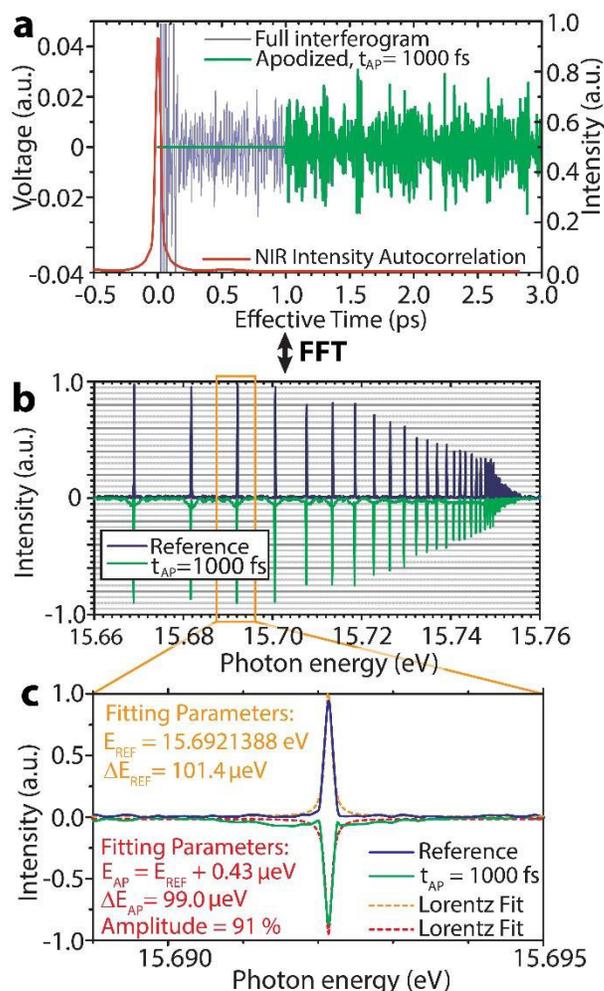

**Fig. 6** Simulation of apodization in order to neglect potentially disturbing parts in the interferogram where the two NIR pulses temporarily overlap. a Blue: full interferogram resulting from an FFT of a simulated argon absorption spectrum retrieved from a convolution of digitized literature data of argon Rydberg states [60] and the 13$^{th}$ harmonic of our laser system, green: apodized interferogram with $t_{AP}$ = 1000 fs. Red: the measured AC trace of our laser system, exhibiting a FWHM pulse duration of 35 fs. b Absorption features corresponding to the full (blue) and apodized (green) interferogram, respectively. c Close-up of the n = 16 Rydberg state (arbitrarily chosen) with Lorentz fits of the respective lines above, for a better comparison. See text for details.

However, this treatment introduces of course distortions to the line shapes that have to be considered and that limit the overall achievable spectral resolution in this configuration. The comparison of distortion effects on the original FTS signal by using different starting times of apodization windows are simulated and can be found in the Supplementary. This procedure qualifies for applications where a spectral resolution on the 10 to 100 GHz (~ 10 – 100 µeV) level is sufficient which applies for most applications in the XUV - except precision spectroscopy.

### 3.3.4    Fast UV signal detection

For many years, not only the generation of UV light was a challenge but also its detection. In the recent years, the technology of UV detectors has advanced such that UV sensitized photo diodes operable down to 10 nm are commercially available. Standard UV sensitive photo multiplier tubes can be used down to 200 nm (i.e. 6.2 eV) with a high bandwidth (~ 500 MHz for the Hamamatsu series).  They have been successfully implemented in several Doppler-free dual comb spectroscopy investigations of rubidium, performed in different laboratories [17,61].

The achievable signal to noise ratio with those detectors can be estimated, for example for a VUV dual comb spectrometer realized by frequency tripling of an Yb fiber laser (~ 343 nm) with the help of the so called quality factor – a useful figure of merit to compare the performance of DCS, also in different spectral ranges [33]. For the operation in the VUV, we calculate the achievable normalized quality factor $M/\sigma_H$ for an acquisition time of 1 s to be ~ $3.2 \cdot 10^5$ Hz$^{1/2}$, with M being the number of spectral elements and $\sigma_H^{-1}$ the signal to noise ratio. The achievable quality factor is about five times lower than standard NIR-DCS can provide, mainly due to the noise of the frequency-tripled lasers and the detector performance. See the supplementary material for the parameters used for the estimation and [33,62] for more details on the quality factor. Signal to noise ratios in the VUV of about half of the ones achieved by state-of-the-art NIR DCS are possible.

Below 200 nm, UV enhanced photodiodes can be implemented. They provide similar response performance as the photomultiplier tubes and fast rise times (~ 1-2 ns, Opto Diode Corp.), allowing detection bandwidths of up to 350 MHz. A balanced circuit that combines the best compromise of high detection sensitivity and sufficient bandwidth (5-40 MHz) is currently under development. We performed a careful estimation regarding the achievable quality factor for the XUV spectral range by implementing "worst case" values for the noise levels in the XUV (see table S1 in the Supplementary). With this, the achievable quality factor in the XUV will be better than $1.2 \cdot 10^4$ Hz$^{1/2}$. Due to the extremely high spectral coverage in the XUV, the expected signal to noise ratio will be two to three orders of magnitude lower than for NIR DCS. Considering the huge absorption cross sections in the UV being up to 5 orders of magnitude larger than in the NIR and reflecting the available signal processing techniques summarized in the next section, XUV DCS holds great potential with adapted measurement times and averaging. With the laser amplifiers and data acquisition systems available, continuous measurements of up to 20 minutes are possible, it should be noted that in contemporary time-domain experiments employing HHG relying on carrier-envelope-phase and interferometer path-length stabilization several hour acquisition times are standard.

*Signal processing*

Given the limited average powers in the UV on the 10-100 µW level and the expected overall elevated noise performance mainly due to the amplifiers, the HHG and detector, we expect that averaging and/or signal processing will be required for achieving decent signal to noise ratios in the UV spectra. For this, mutual coherence between the two lasers has to be achieved. In the meantime, there are several approaches with DCS established:

Servo controls, locking the frequency comb parameters to a frequency standard or locking two lines of the red/blue spectral ends of each comb to narrow-linewidth CW lasers enables a coherence time that is inversely proportional to the linewidth of the CW reference lasers used [63,64]. With this, measurement times on the order of seconds can be achieved and further averaging of those recordings after phase correction can improve the signal to noise ratio further. Recently, feed-forward phase-stable interferometers achieved coherence times of almost 2000 s, both in the NIR and the MIR [65,66].

Another approach is not to control the relative fluctuations but to track them by analogue [67], digital real-time [68] processing or a posteriori corrections [69].

The latest approach is to achieve a certain inherent mutual coherence that comes along with laser cavities that can be operated directly as a dual comb source, either based on dual-wavelength mode-locked lasers [70], bi-directional mode locked lasers [71], counter-propagating microresonators [72] or involving two combs with different polarization on one cavity realized with a birefringent plate [73]. However, using those dual laser sources for UV-DCS will still need further active stabilization or tracking of the fluctuations as individual laser amplifiers for each comb are still required here.

## 4. Conclusion and outlook

In our research article, we presented a roadmap to realize (extreme) ultraviolet dual comb spectroscopy. We proposed a suitable experimental setup for UV-DCS based on cutting-edge technology and performed important studies on its feasibility that open up broadband table-top UV spectroscopy with superior spectral resolution. UV-DCS has the potential to provide a relative resolution on the order of at least $10^{-9}$. This is at least two orders of magnitude better than the best synchrotron beam lines [52,74] while achieving similar brilliance levels like synchrotron beamlines around 20 eV [75]. As a result, studies that so far could only be performed at those facilities and/or with huge time commitment using scanning UV sources can now be intensified with UV-DCS, even beyond the Doppler limit. UV-DCS enables detailed ultra-broadband studies of atomic and molecular gases in a spectral domain that involves element specific valence electron transitions, often superimposed with vibronic excitations but can also address vibrationless Rydberg transitions, all resulting in strong but complex and fine-structured absorption spectra [22,76].

For the implementation of dual comb spectroscopy in the (X)UV, several issues need to be considered, most importantly high-flux UV frequency comb generation, efficient UV interferometry and fast UV signal detection. With state of the art laser technology, we estimate the photon flux sufficient for decent signal to noise ratios up to 30 eV (or 40 nm). We have shown that the noise performance of pulse picking high power frequency combs is mainly dependent on the pulse picking AOM. With a conventional CEO stabilization after the amplifier and stabilization of the optical path, low phase noise measurements resulting in linewidths on the 10 mHz level are possible in the XUV. The repetition rate of the XUV frequency combs should be between 10 and 20 MHz regarding the noise performance but also regarding reasonable dual comb measurement parameters. Beyond that, the amplifier technology and the XUV signal detection will need to be further developed towards higher signal strengths at higher repetition rates.

Regarding efficient XUV interferometry, we have discussed different concepts and identified a promising method for the investigation of long-living (Rydberg) states by NIR superposition and adjusted signal processing.

With all those considerations, we conclude with an estimation of the minimum absorption sensitivity $(\alpha_0 L)_{min}$ at 1 s [33] achievable with DCS for two explicit science cases in the VUV and XUV. For the case of absorption measurements of formaldehyde, centered at 343 nm (3.6 eV), we estimate $(\alpha_0 L)_{min} \sim 5.8 \cdot 10^{-5}$. For the XUV argon Rydberg series around 80 nm (15.6 eV), the achievable minimum absorption sensitivity will be about $(\alpha_0 L)_{min} \sim 4.5 \cdot 10^{-2}$. The potential sensitivity in the XUV will be up to three orders of magnitude lower than in the VUV, mainly due to the lower achievable signal to noise ratio and the largely extended spectral coverage (number of spectral elements). However, when compared to Fourier transform spectrometers aiming for example at differential optical absorption spectroscopy measurements in the VUV, this level of sensitivity is competitive [77]. Our study paves the way for a novel method to answer pressing scientific questions around fundamental processes in the UV like photo-induced non-adiabatic dynamics in molecules but also photolysis and the onset of aerosol nucleation influencing atmospheric reaction path ways.

**Funding.** European Research Council (947288), Deutsche Forschungsgemeinschaft (GRK2101), The Carl Zeiss Foundation, The Daimler and Benz Foundation (32-02/17), Fraunhofer Cluster of Excellence Advanced Photon Sources, This research was funded in whole, or in part, by the Austrian Science Fund (FWF)[Y1254]. For the purpose of open access, the author has applied a CC BY public copyright license to any Author Accepted Manuscript version arising from this submission.

**Acknowledgments.** The authors thank Nathalie Picqué for helpful discussions. Vinzenz Hilbert's help during our first XUV photo diode tests and Maxim Tschernajew`s assistance with the measurements on twin pulse HHG are warmly acknowledged.

**Disclosures.** The authors declare no conflicts of interest.

**Data availability.** Data underlying the results presented in this paper are not publicly available at this time but may be obtained from the authors upon reasonable request.

**Supplemental document.** See **Supplement 1** for supporting content.

# Ultraviolet Dual Comb Spectroscopy: A Roadmap

*(Supplementary Information)*

Vittoria Schuster[1,2], Chang Liu[1,2,3], Robert Klas[1,2,3], Pablo Dominguez[4], Jan Rothhardt[1,2,3,5], Jens Limpert[1,2,3,5], and Birgitta Bernhardt[1,2,6*]

[1]Institute of Applied Physics, Friedrich Schiller University Jena, Albert Einstein Straße 6, 07745 Jena, Germany
[2]Abbe Center of Photonics, Friedrich Schiller University Jena, Albert Einstein Straße 6, 07745 Jena, Germany
[3]Helmholtz-Institute Jena, Fröbelstieg 3, 07743 Jena, Germany
[4]MenloSystems GmbH, Bunsenstraße 5, 82152 Planegg, Germany
[5]Fraunhofer Institute for Applied Optics and Precision Engineering, Albert-Einstein-Straße 7, 07745 Jena, Germany
[6]Intitute of Experimental Physics & Institute of Materials Physics, Graz University of Technology, Petersgasse 16, 8010 Graz, Austria

*Corresponding author: bernhardt@tugraz.at

This document provides supplementary material to the research article "Ultraviolet dual comb spectroscopy: a roadmap".

### A. Phase noise measurement of the fiber amplifiers

For the phase noise measurement, the two laser oscillators have been independently stabilized to a low noise 10 MHz radio frequency reference (ASOPS, Menlosystems). The locking bandwidth used for the measurements was kept low (< 100Hz) allowing to adjust the time overlap of pulses in the optical cross-correlator and showing the free-running behavior of the laser pulses for frequencies above 100 Hz from the carrier. The optical power available for the cross-correlation was set to 10 mW when the pulse picker was not active ($f_{rep}$ = 80 MHz) and was not changed during the measurements. As a result the average power at lower repetition rates (i.e. pulse picker active) was significantly lower, leading also to decreased pulse energies as compared to the 80MHz setting, due to its internal losses. A low pass filter with a cut-off frequency of 250 kHz was placed at the output of the cross-correlator to avoid aliasing effects when performing the Fourier transform of the error signal.

With the amplifier included, a series of phase noise measurements was performed, first with different pump currents to monitor the impact of the pump laser, then with different pulse picking ratios to characterize the influence of the AOM in the amplifier. The results are summarized in figure 3 in the main manuscript, figure S1 presents the integrated phase jitter of the corresponding measurements (in matching color code). Panel 4a shows the phase noise power spectral density for the frequency comb oscillators alone and with one amplifier included with different pump currents corresponding to average output powers of 25 W, 40 W and 70 W, respectively. For all those measurements, the repetition rate was kept constant at 80 MHz. All traces show a broad peak between 400 and 800 Hz that arises from the noisy acoustics of the laboratory environment. They are further increased with the amplifier. Additionally to the general elevated noise floor with the amplifier, the amplifier introduces a distinct noise peak at 30 Hz that is independent of the current of the amplifier pump diode. Several noise spikes above 20 kHz are added by the amplifier, their central frequency and amplitude are dependent of the pump diode, as the inset in panel a shows. The large spike at 160 kHz is originated from the pump-diode drivers of the laser oscillators. It is constant over all measurements. The other spikes originate from the amplifier and/or the pulse-picker.

Panel 4b shows the performance of the amplifier with different pulse picking scenarios, resulting in 80 MHz, 10 MHz and 4 MHz operation, respectively. The overall performance seems to get noisier when the repetition rate is decreased. This is not only arising from the setting of the pulse picking amplifier but also from the concurrent decreased detection sensitivity of the optical cross correlator. For high picking ratios ($f_{rep}$ ≤ 4 MHz), the laboratory acoustics seem to get more salient, and the overall noise floor is increased. In particular, the pulse-picker spike around 20 kHz gets significantly broader. The balanced detection method for single shot measurements will be further improved in order to extend this study also for lower repetition rates down to the kHz level.

### B. Simulation on balanced XUV twin pulse generation

A simulation for the generation of twin XUV pulses via high harmonic generation was carried out with a simple one dimensional model [1]. Here, the number $N_{out}$ of XUV photons emitted on axis per unit of time and area can be assumed to

$$N_{out} \propto \rho^2 A_q^2 \frac{4 L_{abs}^2}{1 + 4\pi^2 \left(\frac{L_{abs}^2}{L_{coh}^2}\right)}.$$

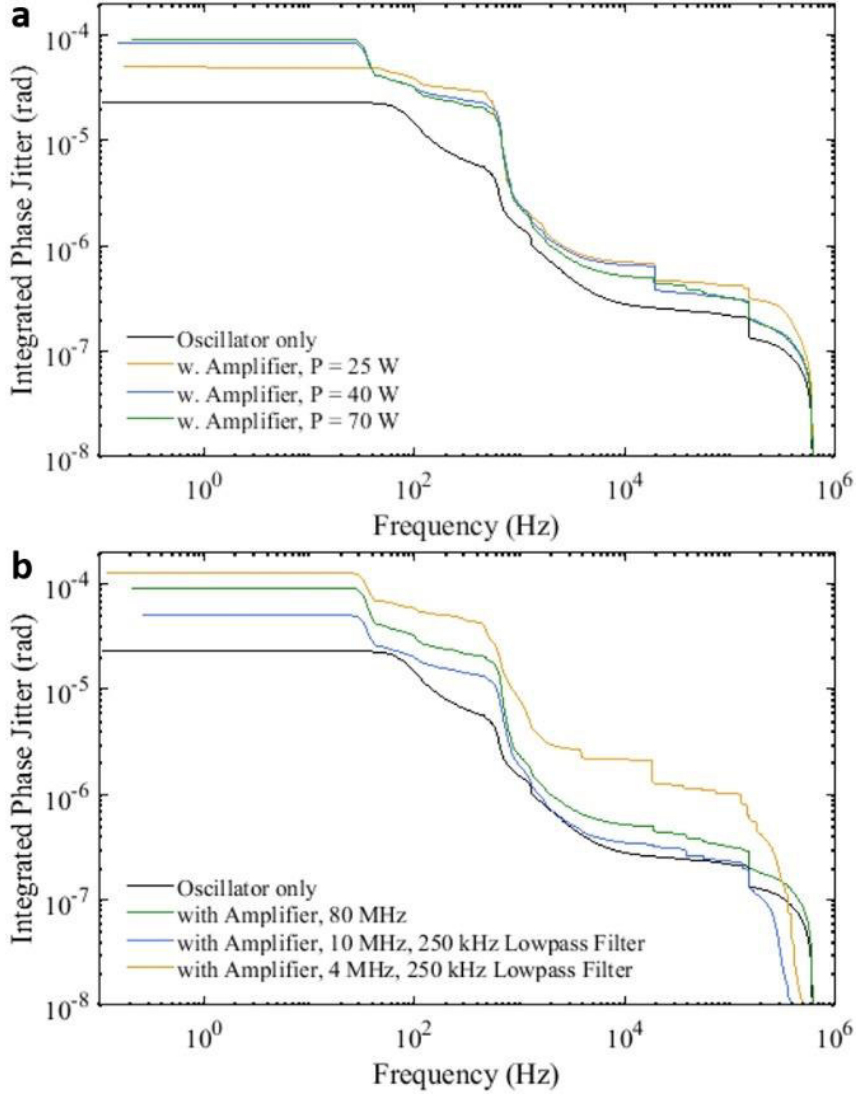

**Figure S1** Integrated Phase Jitter of the two frequency comb oscillators alone (black, depicted in both panels) and with one fiber amplifier included, **a** for different output powers: 25W (orange), 40 W (violet) and 70 W (green), respectively; **b** for different repetition rates: 80 MHz (green), 10 MHz (violet) and 4 MHz (orange).

$$\left[1 + \exp\left(-\frac{L_{med}}{L_{coh}}\right) - 2\cos\left(\frac{\pi L_{med}}{L_{coh}}\right)\exp\left(-\frac{L_{med}}{2L_{abs}}\right)\right],$$

with the medium length $L_{med}$, which in this particular experiment corresponds to the gas jet diameter. The absorption length is calculated as $L_{abs} = 1/\sigma\rho$, where $\rho$ is the gas density and $\sigma$ is the ionization cross section. Further, the coherence length is $L_{coh} = \pi/|\Delta k|$, with $\Delta k$ as the wave vector mismatch between the fundamental and XUV radiation [2].

The amplitude of the atomic dipole response $A_q$ scales as $\left(I/I_{cutoff}\right)^{4.6}$ for $I > I_{cutoff}$, otherwise as $\left(I/I_{cutoff}\right)^{10.6}$, which is almost zero. In this case $I_{cutoff}$ is the minimum intensity required for efficient generation of the given harmonic [3].

The ADK model was used to calculate the ionization of the emitting gas [4]. The driving field was composed of two pulses with a duration of 34 fs each and a separation of 200 fs, which is sufficient for them not to overlap. The ionization of the medium was assumed to remain constant in the time between the two pulses (see fig. S2a).

To calculate the flux generated over the total focus area, the simple one dimensional model was extended to a cylindrically symmetric two dimensional model. For this, the generated photon flux was calculated by integrating over the generated XUV photons in the Gaussian intensity profile of the driving laser.

The XUV flux generated by each of the driving pulses was determined by integrating the effective XUV flux at optimal phase matching pressure over the emission times corresponding to the duration of the respective pulses (see fig. S2b and c). The decreasing XUV signal at higher pressures (fig. S2b) is due to reabsorption by the residual gas in the chamber.
In order to balance the flux of the XUV pulses, the simulation was performed for different ratios of the fundamental pulse energies, such that their sum always amounted to 300 µJ. It was carried out for a fundamental wavelength of 1030 nm. The radius of the driving laser focus was assumed to be 32.5 µm and the diameter of the gas nozzle 65 µm. The results for the 33rd harmonic are represented as solid lines in figure 5b. The simulations predict the flux of the two XUV pulses to be equal if the ratio between the fundamental pulse energies is $E_{P,1}/E_{P,2} \approx 2/3$. This is in good agreement with the experimental results of the pre-study (see 3.3.3) that are represented by circles in Fig. 5b.

### C. Simulation of different apodization windows

For small time delays similar to and smaller than the NIR pulse duration, interference between the two NIR pulse trains occurs, resulting in transient drops and peaks of the XUV output recurrent with a period that is corresponding to the laser's fundamental wavelength. The video available as supplementary material shows this effect while the delay between the two pulses is scanned between + 67 fs to -20 fs. The HHG yield has been optimized before at a fixed delay of 870 fs. The 25th to the 53rd harmonics are shown (~ 30 eV - 64 eV, left side).

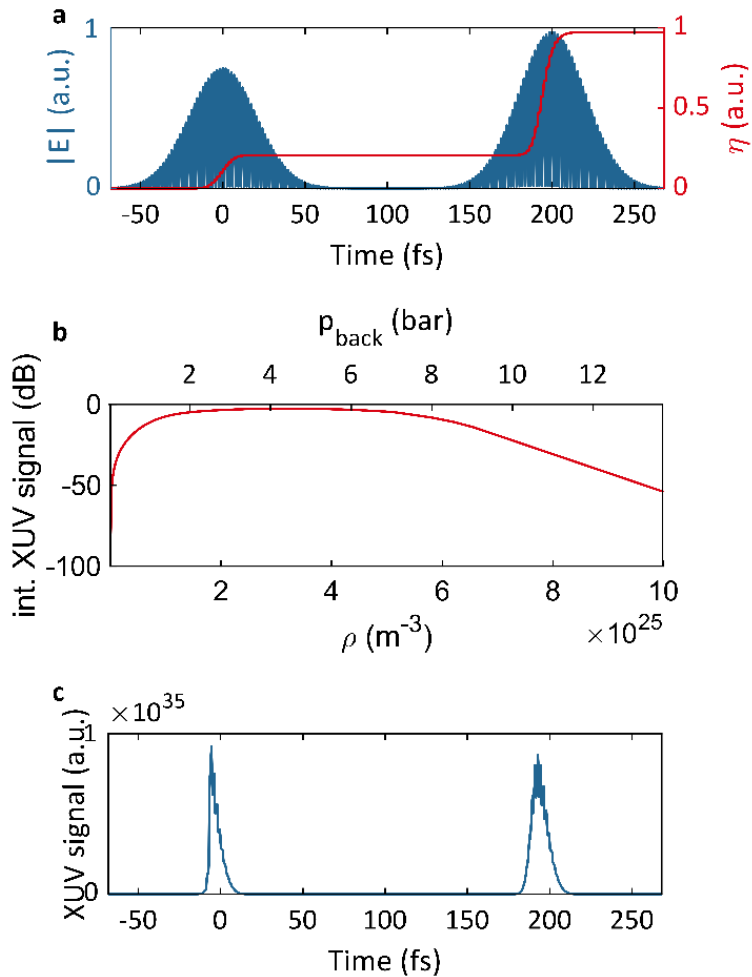

**Figure S2** Simulation on balanced XUV twin pulse generation for the example of fundamental pulse energy of the first pulse 111 µJ and second pulse 189 µJ. a absolute electric field |E| of the fundamental pulses and the ionization η of the emitting gas over time, b determination of the ideal phase matching pressure accounting for reabsorption by residual gas. The evaluated XUV signal is fully integrated over both space and time. c XUV signal at optimal backing pressure of 4.25 bar over the emission time yielding the contribution of the respective XUV pulses to the total signal

To investigate a potential circumvention of this obstructive interference effect, a simulation about the influence of an apodization window onto the retrievable absorption spectrum was performed for the special case of XUV Fourier Transform Spectroscopy of argon Rydberg states converging against the $^2P_{3/2}$ ionization limit. For this, high resolution argon XUV absorption spectra where digitized [5]. Those argon absorption data were adapted to the spectrum of the 13$^{th}$ harmonic of our laser system (see fig S3a). In order to retrieve the corresponding (simulated) interferogram that would arise from an XUV-FTS measurement in argon, the Fourier transform of this spectrum was calculated (fig. S3b). To mimic our experimental conditions, the sampling rate of the retrieved interferogram was set to 2.5 MHz, which corresponding to a mirror scanning speed of 100 mm/s. With the simulation parameters, a spectral resolution on the µeV level in the optical domain was achieved.

By applying apodization windows with different starting times it can be considered how much of the molecular response will be left by analyzing the signal-to-noise ratio of the resulting absorption spectra. Panels c,d,e in figure S3, show different apodization windows with starting times of $t_{AP}$ = 100 fs, 300 fs and 1000 fs, respectively. Additionally, the autocorrelation trace of our NIR amplifier output is shown (red line). Panels f, g, and h show the corresponding remaining absorption features after Fourier transforming the apodized interferogram, respectively. A close-up to one Rydberg state (n=16, arbitrarily chosen) for the three different apodization times shows the influence of the apodization timing on the line shape of the resonances (panels i, j, k for $t_{AP}$ = 100 fs, 300 fs, and 1000 fs, respectively). The line shape parameters change on the µeV level with regard to the line width and on the sub-µeV level with regard to the resonance center positions, even for $t_{AP}$ = 1000 fs. For the latter truncation that even cuts out a small NIR side pulse at 0.5 fs (see autocorrelation trace in fig S3c-e), the amplitude results in 91 % of the incident value. With this, it can be concluded that applying apodization windows to the interferogram is a promising option to neglect potentially disturbing NIR interference effects, provided the NIR pulse duration is small when compared to the sample response time (state lifetimes). However, this treatment introduces of course distortions to the line shapes that have to be considered and that limit the overall achievable spectral resolution in this configuration. For the sake of simplicity we only used rectangular apodization windows that have the effect of building side lobes in the spectrum, as well as the large apodization times for best visibility, resulting in pedestals around the resonances (see figure S3 h). Those side lobes can be decreased with appropriate window shapes like Blackman-Harris, taking the cost of losing some spectral resolution. The shape of window can be chosen by one according to the independent application needs.

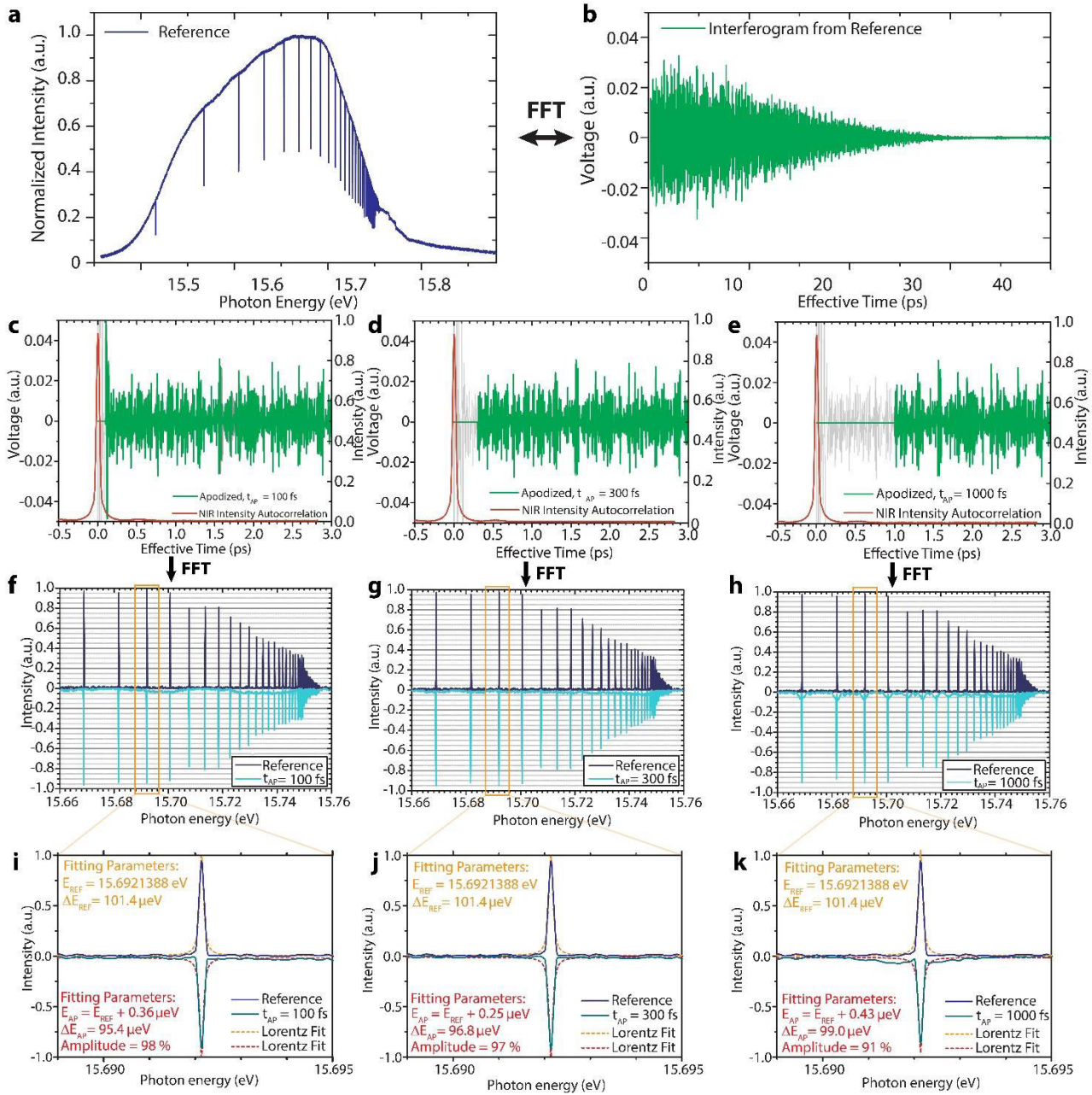

**Figure S3** Simulation of different apodization windows in order to neglect potentially perturbed parts of the interferogram where the two NIR pulses temporarily overlap. **a** The simulated argon absorption spectrum retrieved from a convolution of digitized literature data of argon Rydberg states [5] and the 13[th] harmonic of our laser system **b** corresponding interferogram after Fourier transformation of the spectrum depicted in panel a, **c** Interferogram after applying a rectangular apodization window starting at $t_{AP}$ = 100 fs, **d** $t_{AP}$ = 300 fs and **e** $t_{AP}$ = 1000 fs. The measured AC trace of our laser system is plotted in red exhibiting a FWHM pulse duration of 35 fs. Panels **f**, **g**, **h** show the corresponding absorption features after FFT of the 100 fs, 300 fs and 1000 fs truncations (all in light blue) and, for comparison, the original unapodized spectral features identical with the ones from panel a (dark blue), despite of the missing envelope due to the cut out of central burst by apodization. Panels i, j and k show a close-up of the n = 16 Rydberg state (arbitrarily chosen) with Lorentz fits of the respective lines above. See text for details.

## D.     Estimation of the achievable quality factors and sensitivity in the ultraviolet region

As a useful figure of merit we estimate the achievable normalized quality factor $M/\sigma_H$ at T = 1 s, as it has been done for different NIR dual comb spectrometer configurations in Newbury et al. [6]. M is the number of spectral elements, $1/\sigma_H$ the signal-to-noise ratio and T the acquisition time. We estimate the quality factors for the two science cases in the NUV and in the VUV mentioned in our manuscript, and compare them to one of our earlier Yb-fiber laser based NIR DCS configurations at 1030 nm [7]. Third harmonic generation enables a VUV dual comb spectrometer operating at a central wavelength of 343 nm (3.6 eV) for measurements in formaldehyde and with HHG a DCS operating in the XUV at 80 nm (15.6 eV) is possible for the study of the argon Rydberg series. Table S1 summarizes the parameters that were used for the calculation. We estimate the quality factors $M/\sigma_H$ at T = 1 s to be $1.7 \cdot 10^6$, $3.2 \cdot 10^5$ and $1.2 \cdot 10^4$ for the NIR, the VUV and the XUV, respectively. While the overall performance in the VUV is still promising even for single shot measurements, the quality factor in the XUV dictates further signal processing techniques with averaging. The poorer quality factor in the XUV is mainly due to the noise performance of the amplifiers and the detector, and the huge spectral coverage that is one to two orders of magnitude larger than in the NIR and the VUV, respectively. Spectral filtering with adjusted dual comb parameters could be a possibility to improve the quality factor for first sensitive measurements in the XUV.

Considering absorption cross sections and obtained flux-levels, the signal-to-noise ratio is expected to be about half in the VUV, and up to 3 orders of magnitude lower in the XUV when compared to NIR dual comb spectrometers. With that, high quality spectra from single shot measurements will be possible in the VUV but not in the XUV. However, the absorption cross sections (up to 100 Mb) in the XUV up to five orders of magnitude higher than in the NIR can still result in good signal-to-noise ratios (> 100) in combination with adapted averaging and signal processing schemes.

**Table S1** Definition of the variables used for the estimation of the quality factor of a dual comb spectrometer operating in the NIR, VUV and XUV.

| Quantity (Units) | Variable | NIR [7], 1030 nm or 1.2 eV | VUV, 343 nm or 3.6 eV | XUV, 80 nm or 15.6 eV |
|---|---|---|---|---|
| UV comb power (W) | $P_c$ | $1 \cdot 10^{-3}$ | $1 \cdot 10^{-3}$ | $100 \cdot 10^{-6}$ |
| Spectral width (THz) | $\Delta \nu$ | 7 | 5 | 100 |
| Spectral resolution (GHz) | $\delta f_{opt}$ | 4.5 | 10 | 10 |
| Repetition rate (MHz) | $f_{rep}$ | 128.7 | 80 | 10 |
| Number of spectral elements | $M \equiv \Delta\nu/\delta f_{opt}$ | 1500 | 500 | 10000 |
| Duty cycle | $\epsilon \equiv \delta f_{opt}/f_{rep}$ | 35 | 125 | 1000 |
| Detector Noise (W/Hz$^{1/2}$) | NEP | $2 \cdot 10^{-12}$ | $6 \cdot 10^{-16}$ | $2 \cdot 10^{-17}$ |
| Detection dynamic range | D | 7000 | 7000 | 7000 |
| Laser Relative Intensity Noise (dBc/Hz) | RIN | -145 | -140 | -120 |
| Detector efficiency | $\eta$ | 0.6 | 0.25 | 0.26 |
| # of filters | F | 1 | 1 | 1 |
| # of detectors | $N_d$ | 1 | 1 | 1 |
| **Quality factor at 1 s (Hz$^{1/2}$)** | $M/\sigma_H$ | $1.7 \cdot 10^6$ | $3.2 \cdot 10^5$ | $1.2 \cdot 10^4$ |